\documentstyle[12pt,aaspp4]{article}
\begin{document}
\title{THE POWER SPECTRUM OF CLUSTERS OF GALAXIES AND
THE PRESS-SCHECHTER APPROXIMATION}
\author {MIRT GRAMANN AND IVAN SUHHONENKO}
\affil{Tartu Astrophysical Observatory, T\~oravere EE-2444,
Estonia}

\begin{abstract}
We examine the power spectrum of clusters in the Press-Schechter (PS)
theory and in N-body simulations to see how the power spectrum
of clusters is related to the power spectrum of matter density fluctuations in
the Universe. An analytic model for the power spectrum of clusters for
their given number density is presented, both for real space and 
redshift space. We test this model against results from N-body simulations
and find that the agreement between the analytic theory and the numerical
results is good for wavelengths $\lambda >60h^{-1}$ Mpc. On smaller scales
non-linear processes that are not considered in the linear PS
approximation influence the result. We also use our analytic model to study the
redshift-space power spectrum of clusters in cold dark matter models with
a cosmological constant ($\Lambda$CDM) and with a scale-invariant 
Harrison-Zel'dovich initial spectrum of density fluctuations. We find that 
power spectra of clusters in these models are not consistent with the 
observed power spectra of the APM and Abell-ACO clusters. One possible 
explanation for the observed power spectra of clusters is an inflationary 
scenario with a scalar field with the potential that has a localized 
steplike feature. We use the PS theory to examine the power spectrum of 
clusters in this model.

\end{abstract}

\keywords{methods: numerical --galaxies: clustering -- galaxies:
clusters: general -- cosmology: theory -- large scale structure of universe}

\section{INTRODUCTION}

Clusters of galaxies are efficient tracers of the large-scale structure
of the universe. A strong spatial correlation of clusters of galaxies
(Bahcall \& Soneira 1983; Klypin \& Kopylov 1983) provided some of the
first evidence for the existence of organized structure on large scales.
To date, much effort has been devoted to determine the correlation
function and the power spectrum of clusters of galaxies on large spatial
scales (e.g. Postman, Huchra \& Geller 1992; Peacock \& West 1992;
Einasto et al. 1993; Dalton et al. 1994; Romer et al. 1994;
Croft et al. 1997; Einasto et al. 1997a; Retzlaff et al. 1998; Tadros,
Efstathiou \& Dalton 1998). Figure~1 shows the power spectrum of the
spatial distribution of the Abell-ACO clusters as determined by 
Einasto et al. (1999), and the power spectrum of the APM clusters as 
found by Tadros, Efstathiou \& Dalton (1998). For comparison, we show in 
Figure~1 the power spectrum derived from the distribution of galaxies in 
the APM survey (Baugh \& Efstathiou 1993). Observations give the power 
spectrum of clusters for a given number density of clusters, $n_{cl}$. The 
number density of the APM clusters and Abell clusters is 
$n_{cl} \sim 3.4 \times 10^{-5} h^3$ Mpc$^{-3}$ and
$n_{cl} \sim 2.5 \times 10^{-5} h^3$ Mpc$^{-3}$, respectively
(Dalton et al. 1994; Einasto et al. 1997b; Retzlaff et al. 1998).

As the observational data on the power spectrum of clusters improve,
the need for precise theoretical predictions becomes increasingly important.
The power spectrum of clusters for different cosmological models can be
calculated using N-body simulations (e.g. Retzlaff et al. 1998; Tadros,
Efstathiou \& Dalton 1998; Suhhonenko \& Gramann 1999). However, this
approach is difficult as it requires simulations with a very large dynamical
range to identify correctly a sufficient number of clusters. In this situation
it would be useful to have an analytical approach to describe the
origin of the power spectrum of clusters of galaxies and to predict it.

The correlation function of clusters of galaxies has been a subject
of many attempts at analytical modelling (e.g. Kaiser 1984; Kashlinsky
1987; Cole \& Kaiser 1989; Mann, Heavens, \& Peacock 1993; Mo \& White 1996,
hereafter MW96; Catelan et al. 1998; Kashlinsky 1998). In particular,
using the Press-Schechter (PS) formalism to calculate the correlation
function of clusters in Lagrangian space and mapping from Lagrangian space
to Eulerian space within the context of the spherical collapse model, MW96
have derived an analytical expression for the correlation function of
clusters of mass $M$ as
$$
\xi_{M}(r) = b^2 (M) \, \xi(r),
\eqno(1)
$$
where $\xi(r)$ is the correlation function of matter density fluctuations.
The bias parameter $b(M)$ is
$$
b(M) = 1 + {1 \over \delta_t}
\left({\delta_t^2 \over \sigma^2(M)} -1 \right)  \, ,
\eqno(2)
$$
where $\sigma^2(M)$ is the density dispersion on a mass scale $M$
and $\delta_t=1.686$. This equation has been found to be in good agreement with
N-body results by MW96 and by Mo et al. (1996).

The power spectrum and the correlation function of clusters form a Fourier 
transform pair. Therefore, in the linear approximation the power spectrum of 
clusters of mass $M$ can be expressed as
$$
P_M (k) = b^2(M) \, P(k),
\eqno(3)
$$
where $P(k)$ is the power spectrum of matter density fluctuations.

In this paper we develop further the results obtained by MW96 and
calculate the power spectrum and the correlation function
of galaxy clusters for a given number density of clusters $n_{cl}$
in Lagrangian space and in Eulerian space (eq. [15,23-24] and [25-27] below).
Observations provide the distribution of clusters in redshift
space, which is distorted due to peculiar velocities of clusters.
In order to study the power spectrum of clusters in redshift space,
we use the linear approximation derived by Kaiser (1987). In \S 3 we
examine the cluster power spectrum in real space and in redshift space
using N-body simulations with realistic initial power spectra and show
that in both cases the PS theory gives an accurate description of the power
spectrum of clusters at wavenumbers $k<0.1h$ Mpc$^{-1}$ (or at
wavelengths $\lambda >60h^{-1}$ Mpc). In \S 4 we use our model to
study the redshift-space power spectrum of clusters in different
cold dark matter models. We study inflationary models with a
scale-invariant initial spectrum of density fluctuations and with a
steplike initial spectrum derived by Starobinsky (1992).
\S 5 summarizes the main results.

A Hubble constant of $H_0=100h$ km s$^{-1}$ Mpc$^{-1}$ is used
throughout this paper.

\section{THE POWER SPECTRUM OF CLUSTERS IN THE PS APPROXIMATION}

An important aspect of the PS model is that, being entirely based on
linear theory, suitably extrapolated to the collapse time of spherical
perturbations, it is, by definition, local in Lagrangian space. While
this Lagrangian aspect of the theory does not have immediate
implications for the study of the mass function of clusters of
galaxies, it is important for the study of their spatial clustering
properties.

Let us assume that the density contrast $\epsilon_M({\bf q})$
on a mass scale $M$ is a Gaussian random field. Here ${\bf q}$ represents
the comoving Lagrangian coordinate. The probability that
one would find a density contrast between $\epsilon_M$ and
$\epsilon_M +d\epsilon_M$ is
$$
p(\epsilon_M) \, d\epsilon_M = {1 \over \sqrt{2\pi} \sigma(M} \,\,
\exp \left[- {\epsilon_M^2 \over 2\sigma^2(M)}\right] \,\, d\epsilon_M \, .
\eqno (4)
$$
The density dispersion $\sigma^2(M)$ can be written as
$$
\sigma^2(M) \equiv < \epsilon^2_M >=
{1 \over 2 \pi^2} \int P(k) \, W^2 \,(kR)\, k^2 \, dk,
\eqno(5)
$$
where $W(kR)$ is the Fourier transform of the window function applied to
determine the density field. In this paper we will use the top-hat window
function. For the top-hat window, the mass $M$ is related to the window
radius $R$ as $M=4\pi\rho_b R^3/3$ (where $\rho_b$ is the mean
background density).

Press \& Schechter (1974) suggested that we can determine the probability
that a cluster of mass $M$ has formed as
$$
\Pi_M = {\int_{\delta_t} ^{\infty}} \, p(\epsilon_M) \, d\epsilon_M
= {1 \over 2}\, \, {\rm erfc} \left[\delta_t \over \sqrt 2 \sigma(M)\right] \, ,
\eqno(6)
$$
where $\delta_t$ is a certain threshold value. PS used a value
$\delta_t=1.686$, which is motivated by the spherical collapse model.
The number density of clusters with the mass between $M$
and $M+dM$ is given by (Press \& Schechter 1974)
$$
n(M) dM = 2 \,\, {\rho_b \over M} \,\, {d\Pi_M \over dM} \,\, dM =
- \sqrt{{2 \over \pi}} {\rho_b \over M} \,{\delta_t \over \sigma^2(M)} \,
{d\sigma(M)\over dM} \,\,
\exp \left[-{\delta_t^2 \over 2\sigma^2(M)}\right] \, dM \,.
\eqno(7)
$$
Equation (7) involves an additional correction factor 2 to allow all
the matter in the universe to form bound structures (see Peacock \&
Heavens 1990; Bond et al. 1991 for a more detailed discussion of
this correction). Therefore, the number density of clusters of mass
larger than $M$ can be expressed as
$$
n_{cl}(>M) = \, {\int_M^{\infty}} n(M') dM' =
 - {3 \over (2\pi)^{3/2}} {\int_R^{\infty}} { \delta_t \over \sigma^2(r)} \,
{d\sigma(r)\over dr} \,\,
\exp\left[-{\delta_t^2 \over 2\sigma^2(r)}\right] \, \,\,
{dr \over r^3} \, .
\eqno(8)
$$
Equation (8) has been frequently used to determine the cluster
abundance in different cosmological models (e.g. Efstathiou et al.
1988; White, Efstathiou \& Frenk 1993; Lacey \& Cole 1994; 
Eke, Cole \& Frenk 1996; Borgani et al. 1997).

For a two-dimensional Gaussian distribution the probability density for
finding simultaneously $\epsilon_{M_1}$ on scale $M_1$ and
$\epsilon_{M_2}$ on scale $M_2$ separated by $q$ is given by
$$
p(\epsilon_{M_1},\epsilon_{M_2}) =
{1 \over 2 \pi \sigma(M_1) \sigma(M_2) \sqrt{1 - \rho^2_{12}}} \,\,
{\exp\left[-{ 1 \over 2 (1-\rho^2_{12})} (x_1^2 - 2 \rho_{12} x_1 x_2 +
x_2^2)\right]}  ,
\eqno(9)
$$
where $x_1=\epsilon_{M_1}/\sigma(M_1)$, $x_2=\epsilon_{M_2}/\sigma(M_2)$
and $\rho_{12} = \xi_{12}(q)/\sigma(M_1)\sigma(M_2)$ is the correlation
coefficient between $\epsilon_{M_1}$ and $\epsilon_{M_2}$. The
$\xi_{12}(q)$ is the two-point correlation function of the linear
density contrast smoothed on the scales $M_1$ and $M_2$:
$$
\xi_{12}(q) \equiv <\epsilon_{M_1}({\bf q_1}) \epsilon_{M_2}({\bf
q_2})> = {1 \over 2 \pi^2} \int P(k) \, W (kR_1) \, W(kR_2) j_0 (kq)
k^2 \, dk,
\eqno(10)
$$
where $q=\vert {\bf q_1} - {\bf q_2} \vert$ and $j_0(x)$ is the
spherical Bessel function of the order zero. Similarly to equation (6), we
can write for the probability for two clusters of mass $M_1$, $M_2$ to
form as
$$
\Pi_{M_1 M_2} = {\int_{\delta_t} ^{\infty}}
{\int_{\delta_t} ^{\infty}} \, p(\epsilon_{M_1},\epsilon_{M_2}) \,
d\epsilon_{M_1} d\epsilon_{M_2} \, \, .
\eqno(11)
$$
To determine the correlation function of clusters of masses
[$M_1$;$M_1+dM_1$] and [$M_2$;$M_2+dM_2$] we can use an expression
$$
1 + \xi_{M_1 M_2} (q) = {{\partial^2 \Pi_{M_1 M_2} \over
\partial M_1 \partial M_2}
\over {d\Pi_{M_1} \over dM_1} {d\Pi_{M_2} \over dM_2}}.
\eqno(12)
$$
A similar equation was used by Kashlinky (1987, 1998) to derive the correlation
function of clusters in Eulerian space and by Catelan et al. (1998) to
describe clustering in Lagrangian space.

In order to find the correlation function of clusters of mass larger
than $M$, we start from approximations (6-7) and (11-12) and write down
the probability to find two clusters in small volumes $dV_1$, $dV_2$,
one cluster being in the mass range [$M_1$, $M_1+dM_1$],
the other in [$M_2$, $M_2+dM_2$], as
$$
d\Phi (M_1,M_2)=4 \, {\rho^2_b \over M_1 M_2} \,\,
{\partial^2 \Pi_{M_1 M_2} \over \partial M_1 \partial M_2} \,\,
dM_1 dM_2 dV_1 dV_2 \, .
\eqno (13)
$$
We included in equation (13) a correction factor 4 to make it
consistent with equation (7) at $\xi_{12}(q)=0$. The probability to find two
clusters of mass larger than $M$ can be expressed as
$$
d\Phi_{cl}(>M)= {\int_{M} ^{\infty}} {\int_{M} ^{\infty}} d\Phi
(M_1,M_2) dM_1 dM_2  = n_{cl}^2 [1 + \xi_{cl}(q)] \, dV_1 dV_2 ,
\eqno (14)
$$
where $n_{cl}$ is the number density of clusters of mass larger
than $M$ (equation 8). Therefore, the correlation function of clusters
can be written in the form
$$
1 + \xi_{cl}(q) ={1 \over n_{cl}^2} \,\, {\int_{M} ^{\infty}}
{\int_{M} ^{\infty}} \, 4 \, {\rho^2_b \over M_1 M_2} \,\,
{\partial^2 \Pi_{M_1 M_2} \over \partial M_1 \partial M_2} \,\, dM_1 dM_2.
\eqno(15)
$$
Equations (5,8-11) and (15) determine the correlation function of clusters
for a given $n_{cl}$ in the PS theory of gravitational clustering.

Let us consider the correlation function of clusters in the linear
approximation, i.e. neglecting the terms involving $\xi^2_{12}(q)$ in
equation (9). In the linear approximation, the probability density
$$
p(\epsilon_{M_1},\epsilon_{M_2}) =
{1 \over 2 \pi \sigma(M_1) \sigma(M_2)} \,
\left[1 + {\xi_{12}(q) \epsilon_{M_1} \epsilon_{M_2}
\over \sigma^2(M_1)\sigma^2(M_2)}\right]
\exp \left[- {\epsilon_{M_1}^2 \over 2\sigma^2(M_1)}\right] \,
\exp \left[- {\epsilon_{M_2}^2 \over 2\sigma^2(M_2)}\right] .
\eqno(16)
$$
Substituting approximation (16) into equation (11) we find that
$$
\Pi_{M_1 M_2} = \Pi_{M_1} \Pi_{M_2} + {1 \over 2 \pi} {\xi_{12}(q) \over
\sigma(M_1) \sigma(M_2)} \,
\exp \left[- {\delta_t^2 \over 2\sigma^2(M_1)}\right] \,
\exp \left[- {\delta_t^2 \over 2\sigma^2(M_2)}\right]
\eqno (17)
$$
and
$$
{\partial^2 \Pi_{M_1 M_2} \over \partial M_1 \partial M_2} =
{d\Pi_{M_1} \over dM_1} \, {d\Pi_{M_2} \over dM_2} \,\,
\left[1 \, + \, {\xi_{12}(q) \over \delta_t^2}
\left({\delta_t^2 \over \sigma^2(M_1)} -1 \right)
\left({\delta_t^2 \over \sigma^2(M_2)} -1 \right)\right] .
\eqno(18)
$$
For separations $q$ much larger that the smoothing lengths, $q >>R_1$ and
$q>>R_2$, the influence of the window functions on the correlations is
negligible, and the correlation function $\xi_{12} (q) \simeq \xi (q)$
(where $\xi(q)$ is the linear mass autocorrelation function). Therefore, in
the linear approximation the correlation function of clusters of masses
[$M_1$;$M_1+dM_1$] and [$M_2$;$M_2+dM_2$] can be written as
$$
\xi_{M_1 M_2} (q) = b^L (M_1) \, b^L (M_2) \xi (q),
\eqno(19)
$$
where the Lagrangian bias factor $b^L(M)$ is
$$
b^L (M) = {1 \over \delta_t} \left({\delta_t^2 \over \sigma^2(M)}
-1 \right)  \, .
\eqno (20)
$$
MW96 derived this equation to describe clustering in Lagrangian
space starting from the behaviour of the conditional Lagrangian mass
function derived by Bond et al. (1991). By using the spherical collapse
model, MW96 found that in Eulerian space the correlation function of clusters
of masses [$M_1$;$M_1+dM_1$] and [$M_2$;$M_2+dM_2$] can be written as
$$
\xi_{M_1 M_2} (r) = b (M_1) \, b (M_2) \xi (r),
\eqno(21)
$$
were $\xi(r)$ is the correlation function of matter density
fluctuations in Eulerian space and the Eulerian bias factor $b (M)$ is
$$
b(M)= 1 + b^L (M) = 1+ {1 \over \delta_t} \left({\delta_t^2 \over \sigma^2(M)}
-1 \right)  \, .
\eqno (22)
$$
Catelan et al. (1998) used a more general method for evolving
the spatial distribution of clusters and found a similar result for
the linear approximation. The shift by 1 of the linear bias factor,
caused here by the transformation from the Lagrangian to the Eulerian
world, comes from mass conservation in Eulerian space.

Let us consider the correlation function of clusters for a given number
density in the linear PS approximation. Substituting approximation (18) into
equation (15) we find that in the linear approximation the Lagrangian
correlation function of clusters for a given $n_{cl}$ can be expressed as
$$
\xi_{cl}(q)=(b_{cl}^L)^2 \, \xi (q)\,,
\eqno(23)
$$
where the bias parameter $b_{cl}^L$ can be written in the form
$$
b_{cl}^L= {1 \over n_{cl}} {\int_M^{\infty}} b^L(M') n(M') dM' =
- {3 \over (2\pi)^{3/2}n_{cl}} {\int_R^{\infty}}
{ 1 \over \sigma^2(r)} \, {d\sigma(r)\over dr}
\left[{\delta_t^2 \over \sigma^2(r)} -1 \right]
\exp\left[-{\delta_t^2 \over 2\sigma^2(r)}\right]
{dr \over r^3} .
\eqno(24)
$$
Similarly to equations (21-22) the Eulerian correlation function of
clusters for a given number density $n_{cl}$ can be written as
$$
\xi_{cl}(r)=b_{cl}^2 \, \xi (r)\,,
\eqno(25)
$$
where the bias parameter $b_{cl}$ is
$$
b_{cl}= 1+ b_{cl}^L= 1 - {3 \over (2\pi)^{3/2}n_{cl}} {\int_R^{\infty}}
{ 1 \over \sigma^2(r)} \, {d\sigma(r)\over dr}
\left[{\delta_t^2 \over \sigma^2(r)} -1 \right]
\exp\left[-{\delta_t^2 \over 2\sigma^2(r)}\right]
{dr \over r^3} .
\eqno(26)
$$
The power spectrum of clusters for a given number density
in the linear approximation can be expressed as
$$
P_{cl}(k)=b_{cl}^2 \, P(k).
\eqno(27)
$$
The cluster bias parameter $b_{cl}$ depends on the minimal mass $M$ (or
the window radius $R$) of clusters and on the power spectrum of density
fluctuations, $P(k)$, which determines the function $\sigma(r)$. For a
fixed $P(k)$ and $n_{cl}$, the minimal mass $M$ (or scale $R$) can be
determined by inverting equation (8). In this approach the power spectrum
(or the correlation function) of clusters does not depend on the
mean background density $\rho_b$ (or the density parameter $\Omega_0$).

Observations provide the distribution of clusters in redshift
space, which is distorted due to peculiar velocities of clusters. On
large scales, where linear theory applies, the power spectrum of matter
density fluctuations in redshift space is given by (Kaiser 1987):
$$
P^s(k)=\left[1 + {2 f(\Omega_0) \over 3} + {f^2(\Omega_0)
\over 5} \right] P(k),
\eqno(28)
$$
where $f(\Omega_0)\approx \Omega_0^{0.6}$ is the linear
velocity growth factor. In the linear approximation (27), relation (28)
takes the form
$$
P^s_{cl}(k)=
\left[1 + {2 f(\Omega_0) \over 3 b_{cl}} + {f^2(\Omega_0) \over
5 b_{cl}^2} \right] b_{cl}^2 \, P(k).
\eqno(29)
$$
Equation (29) determines the power spectrum of clusters
for a given $n_{cl}$ in redshift space.

\section{NUMERICAL RESULTS}

For testing the solutions (26-27) and (29), we ran N-body simulations,
using a particle-mesh code described by Gramann (1988). We investigated
the evolution of $256^3$ particles on a $256^3$ grid, with $\Omega=1$.
The comoving box size was $L=384h^{-1}$ Mpc. The initial density field
was taken to be Gaussian.

We examined the distribution of clusters in two cosmological models
which start from the observed power spectra of the distribution of
galaxies and clusters of galaxies. In the model (1), the initial
linear power spectrum of density fluctuations was chosen in the form
$P(k) \propto k^{-2}$ at wavelengths $\lambda<120h^{-1}$ Mpc. In the
model (2), we assumed that the initial power spectrum contains a
primordial feature at the wavelengths $\lambda \sim 30-60h^{-1}$ Mpc
that correspond to the scale of superclusters of galaxies.
Suhhonenko \& Gramann (1998) studied the mass function, peculiar
velocities, the power spectrum and the correlation function of clusters
in both models for different values of the density parameter $\Omega_0$ and
$\sigma_8$ (the rms fluctuation on the $8h^{-1}$ Mpc scale). The
results were compared with observations. They found that in many aspects
the initial power spectrum of density fluctuations in the model (2) fits the
observed data better than the simple power law model (1).

Clusters were selected in the simulations as maxima of the density
field that was determined on a $256^3$ grid using the CIC-scheme. To
determine peculiar velocities of clusters, we determined the peculiar
velocity field on a $256^3$ grid using the CIC-scheme and found the
peculiar velocties at the grid points were clusters had been
identified. The clusters were then ranked according to their density and we
selected $N_{cl}=(L/d_{cl})^3$ highest ranked clusters to
produce cluster catalogs with a mean intercluster separation
$d_{cl}=30 - 40h^{-1}$Mpc. For comparison, the mean
separation of the observed APM and Abell clusters is $d_{cl} \sim 31h^{-1}$
and $d_{cl} \sim 34h^{-1}$ Mpc, respectively (Dalton et al. 1994;
Retzlaff et al. 1998).

It is difficult to follow the evolution of rich clusters by using
N-body simulations, as it requires simulations with a very large
dynamical range to identify correctly a sufficient number of clusters.
We determined clusters as maxima of the density field smoothed
on the scale $R \sim 1.5h^{-1}$ Mpc. This method of identifying
clusters is not really identical to that what the PS theory is predicting -
the location of collapsed, virialized objects - and this could affect the
degree of agreement between the analytic theory and the numerical results.
However, in order to increase the resolution of simulations we must
increase the number of test particles and grid points, or we have to follow
the evolution of clusters in a smaller box. While the first possibility is
technically difficult, in the latter case the number of rich clusters
becomes too small to get statistically reliable results. Taking into
account the requirements on the number of clusters and on the resolution
together with the fact of fixed computer recources we decided to use a box
size $L=384h^{-1}$ Mpc and a grid size $R_g=1.5h^{-1}$ Mpc.

Figure~2 shows the power spectrum of clusters with a mean separation
$d_{cl}=30h^{-1}$ Mpc in our models. We studied also the power spectrum of
clusters with a mean separation $d_{cl}=35h^{-1}$Mpc and
$d_{cl}=40h^{-1}$ Mpc, and found similar results. To calculate the power
spectrum of clusters in the simulations we determined the density field
of clusters on a $128^3$ grid using the CIC-scheme and calculated its Fourier
components, subtracting the shot noise term. We determined also the Poisson
error bars for the power spectrum. In a shell of $k$-space containing $m$
modes, the Poisson error can be estimated
as $\Delta P_{cl} (k) = 2^{1/2} \, m^{-1/2} d_{cl}^{\,3}$ (see e.g.
Peacock \& Nicholson (1991) for a more detailed analysis of the Poisson
errors in the power spectrum).

Figure~2a shows the results for the model (1) and Figure~2b for the
model (2). For the model (1), the clusters were determined in the simulation
for $\sigma_8=0.5$ and $\sigma_8=0.8$. For the model (2), these parameters were
$\sigma_8=0.5$ and $\sigma_8=0.84$. For comparison, we examined
the power spectrum of clusters in the linear PS approximation (27). We
inverted equation (8) to determine the minimal radius $R$ for a given
$n_{cl}$ (or $d_{cl}$) and after that we calculated the bias factor
$b_{cl}$ using equation (26). For a given $\sigma_8$ and $n_{cl}$, the
bias parameter $b_{cl}$ in the model (1) is similar to that in the model
(2). For $\sigma_8=0.5$, we find that $b_{cl}= 4.3$ and $4.9$ for
separations $d_{cl}=30$ and $40h^{-1}$Mpc, respectively. For $\sigma_8=0.8$,
$b_{cl}=2.9$ and $3.3$, respectively.

Now we can compare the numerical results and the PS theory predictions.
First, consider the power spectrum of clusters at wavenumbers
$k=0.04-0.1h$ Mpc$^{-1}$ (or at wavelengths $\lambda= 60-160 h^{-1}$ Mpc).
On larger scales, the number of modes in the simulations is too small to
get statistically reliable results. Figure~2 demonstrates that at
wavenumbers $k=0.04-0.1h$ Mpc$^{-1}$, the power spectrum of clusters
in the simulations is linearly enhanced with respect to the power spectrum
of the matter distribution in both models studied. For $\sigma_8=0.5$, the
agreement between the results of the N-body simulations and the PS theory
predictions is very good. The mean deviation between the numerical results
and the theoretical predictions is about $4$\% and $7$\% in the model (1)
and in the model (2), respectively. For $\sigma_8 \approx 0.8$, we find
that the power spectrum of clusters in N-body simulations is somewhat lower
than predicted by approximation (27). The power spectrum of clusters is
about $80$\% and $70$\% of the linear theory predictions in the
model (1) and model (2), respectively. The PS approximation predicts
that during the evolution between the $\sigma_8=0.5$ and $\sigma_8=0.8$,
$P_{cl}(k)$ increases slightly. In simulations we find that during the
evolution the power spectrum of clusters decreases. This effect is probably
caused by merging of very rich clusters. Further study (e.g. numerical
simulations with a higher dynamical range) is needed to determine whether
this is a real effect in our models, or a numerical effect due to the
limited dynamical range of the N-body simulations.

Let us now consider the power spectrum of clusters at wavenumbers
$k=0.1-0.2h$ Mpc$^{-1}$ ($\lambda=30-60h^{-1}$ Mpc). In the model (1),
the numerical results on these scales are in good agreement with the
linear approximation (27). But in the model (2), the power spectrum of
clusters in the simulation is significantly smaller than that predicted by the
linear theory. For $\sigma_8=0.84$, we find that the power spectrum of
clusters is only $30$\% of the linear theory predictions. Therefore, in the
model (2), we cannot use the linear approximation to study the power
spectrum of clusters at wavenumbers $k>0.1h$ Mpc$^{-1}$. The power
spectrum of clusters on these scales is probably determined by non-linear
processes in superclusters of galaxies.

Figure~3 shows the redshift-space power spectrum of clusters in the
simulations and in approximation (29). For clarity, we did not plot
error bars in this Figure. In order to study  peculiar velocities
of galaxy clusters and their distribution in redshift space in
models with different $\Omega_0$, we determined peculiar velocities of
clusters in the simulations with $\Omega=1$ and assumed that peculiar
velocities of clusters are proportional to the linear growth factor
$f(\Omega_0)$. Let us consider the redshift-space power spectrum of 
clusters at wavenumbers $k=0.04-0.1h$ Mpc$^{-1}$, where we expect that 
the linear approximation (29) applies. Numerical results show that on 
these scales the power spectrum of clusters in redshift space is linearly 
enhanced with respect to the power spectrum of clusters in real space. 
In Figure~3 we show the power spectrum of clusters in redshift space for 
$\sigma_8=0.5$ and $\Omega=1$. In this case, the agreement between the 
numerical results and approximation (29) is very good. The mean deviation 
between the numerical results and theoretical predictions is about $5$\% and 
$7$\%, in the model (1) and model (2), respectively. For 
$\sigma_8 \approx 0.8$ we find that, similarly to 
real space, the redshift-space power spectrum of clusters in the
simulations is somewhat lower than that predicted by the linear theory.
For $\Omega=1$, the power spectrum of clusters is about $80$\% and
$75$\% of the linear theory predictions in the model (1) and
model (2), respectively. For $\Omega_0=0.2$, we found similar
results. Within the uncertainties due to various numerical
effects in the simulations, the agreement between the analytic theory
and the numerical results is good. The degree of agreement is similar
in real and in redshift space.

\section{MODELS WITH COLD DARK MATTER}

Now we use approximation (29) to analyze the redshift-space power spectrum
of clusters in different cold dark matter models. We examine flat 
cosmological models with the density parameter $\Omega_0=0.3-0.4$, the 
baryonic density $\Omega_B=0.015h^{-2}$ and the normalized Hubble constant 
$h=0.5-0.7$. These parameters are in agreement with recent nucleosynthesis 
results (Tytler at al. 1996), with measurements of the density parameter
(e.g. Dekel, Burstein \& White 1996; Bahcall \& Fan 1998) and with
measurements of the Hubble constant using various distance indicators
(e.g. Tammann 1998). To restore the spatial flatness in the low-density
models, we assume a contribution from a cosmological constant:
$\Omega_{\Lambda}=1-\Omega_0$.

Figure~4 shows the redshift-space power spectrum of clusters in the
$\Lambda$CDM models with a scale-invariant Harrison-Zel'dovich initial
spectrum of density fluctuations ($P_{in}(k) \propto k$). We have used
the transfer function derived by Bardeen et al. (1986) and Sugiyama (1995),
and the COBE normalization derived by Bunn \& White (1997). Figure~4a
demonstrates the power spectrum of the model clusters and the Abell-ACO 
clusters with a mean separation $d_{cl}=34h^{-1}$ Mpc. We show the
power spectrum of the Abell-ACO clusters as found by Einasto et al. (1999).
This spectrum represents the weighted mean of the power spectra 
determined by Einasto et al. (1997a) and Retzlaff et al. (1998). 
Einasto et al. (1997a) determined the power spectrum of the Abell-ACO 
clusters from the correlation function of clusters, while Retzlaff et al. 
(1998) estimated the power spectrum directly (see Einasto et al. 1999 
for details). The power spectrum of the distribution of the Abell-ACO 
clusters peaks at the wavenumber $k=0.052h$ Mpc$^{-1}$ (or at the 
wavelength $\lambda=120h^{-1}$ Mpc). For $k>0.052h$ Mpc$^{-1}$, the cluster 
power spectrum is well approximated by a power law, $P(k) \propto k^n$, 
with $n \approx -1.9$. A similar peak in the one-dimensional power 
spectrum of a deep pencil-beam survey was detected by Broadhurst et al. 
(1990) and in the two-dimensional power spectrum of the Las Campanas 
redshift survey by Landy et al. (1996). Figure~4b shows the power spectrum 
of the model clusters with a mean separation $d_{cl}=31h^{-1}$ Mpc. For 
comparison, we show the power spectrum of the observed APM clusters 
determined by Tadros, Efstathiou \& Dalton (1998). They analyzed the 
redshift survey of 364 clusters described by Dalton et al.
(1994). The mean intercluster separation of APM clusters is
$d_{cl} \sim 31h^{-1}$ Mpc (Dalton et al. 1994). 

Figure~4 shows that the power spectrum of clusters predicted in the 
$\Lambda$CDM models is not consistent with the observed spectra of 
clusters. We studied a $\chi^2$ probability at wavenumbers 
$k=0.03-0.1h$ Mpc$^{-1}$, where we expect that the linear approximation 
(29) applies. For the models presented in Figure~4, the probability to 
fit the observed power spectra of the Abell and APM clusters is less than 
$5\times 10^{-2}$ and $10^{-6}$, respectively. We also studied a 
$\chi^2$ probability for the models, where the amplitude of the power 
spectrum of clusters is about $70$\% and $80$\% of the linear theory 
predictions and found similar results. The power spectrum of clusters 
in the $\Lambda$CDM models with a scale-invariant initial spectrum is 
not consistent with the observed spectra of the APM and Abell-ACO clusters 
(see also e.g. Tadros, Efstathiou \& Dalton 1998; Einasto et al. 1997a).

The power spectrum of density fluctuations in the universe depends on
the physical processes in the early universe. The peak in the power
spectrum of clusters at wavelength $\lambda \simeq 120h^{-1}$ Mpc may
be generated during the era of radiation domination or earlier.
Baryonic acoustic oscillations in adiabatic models may explain the
observed power spectrum only if currently favored determinations of
cosmological parameters are in substantial error (e.g. the density
parameter $\Omega_0<0.2h$) (Eisenstein et al. 1998). One possible
explanation for the observed power spectra of clusters is an inflationary
model with a scalar field whose potential $V(\varphi)$ has a local steplike
feature in the first derivative. This feature can be produced by fast
phase transition in physical field other than an inflaton field. 
An exact analytical expression for the scalar (density) perturbations 
generated in this inflationary model was found by Starobinsky (1992) 
(see also Lesgourgues, Polarski \& Starobinsky 1998). The initial power 
spectrum of density fluctuations in this model can be expressed as
$$
P_{in}(k) \propto {k \, S(k/k_0,p) \over p^2},
\eqno(30)
$$
where function $S(k/k_0,p)$ can be written in the form
$$
S(y,p) =1 -
{3 \, p_1 \over y}\left[f_1(y)\sin{2y} + {2 \over y} \cos{2y}\right] +
{9 \, p_1^2 \, f_2(y) \over 2 \, y^2}
       \left[f_2(y)+f_1(y)\cos{2y} - {2 \over y} \sin{2y} \right] \, .
\eqno(31)
$$
Here, function $f_1(y)=1-y^{-2}$, $f_2(y)=1+y^{-2}$ and $p_1=p-1$.
The initial power spectrum in this model depends on two parameters
$k_0$ and $p$. The parameter $k_0$ determines the location of the step
and the parameter $p$ - the shape of the initial spectrum.
For $p=1$, we recover the scale-invariant Harrison-Zel'dovich
spectrum ($S(y,1) \equiv 1$). At present, the initial spectrum (30-31)
is probably the only example of a initial power spectrum with the
desired properties, for which a closed analytical form exists.

Figure~5 shows the power spectrum of clusters in the $\Lambda$CDM models
with a steplike initial power spectrum. As in Figure~4, we have used
the transfer function derived by Bardeen et al. (1986) and Sugiyama (1995),
and the COBE normalization derived by Bunn \& White (1997). For the
models with $p<1$ and $p>1$, the step parameter was chosen to be
$k_0=0.016h$ Mpc$^{-1}$ and $k_0=0.03h$ Mpc$^{-1}$, respectively. In this
case, in the models with $p<1$, the power spectrum has a well-defined maximum
at the wavenumber $k\simeq 0.05h$ Mpc$^{-1}$ and a second maximum at
$k \simeq 0.1h$ Mpc$^{-1}$. In the models with $p>1$, the picture is inverted.
The power spectrum has a flat upper plateau at wavenumbers
$k<0.05h$ Mpc$^{-1}$, a sharp decrease on smaller scales
($k=0.05-0.1h$ Mpc$^{-1}$) and a secondary maximum at
$k \simeq 0.15h$ Mpc$^{-1}$. At wavenumbers $k>0.05h$ Mpc$^{-1}$,
the power spectrum in the $\Lambda$CDM models with $p \simeq 1.3-1.4$ 
is similar to the power spectrum in the numerical model (2), that we 
studied in \S 3 (see Figure~2b and Figure~3b).

Figure~5a shows the power spectrum of the model clusters and the Abell clusters
with a mean separation $d_{cl}=34h^{-1}$ Mpc. We have studied
inflationary models with parameter $p=0.6-0.8$. Figure~5a shows that the 
shape of the cluster power spectrum in these models is in good 
agreement with the observed power spectrum of the Abell-ACO clusters.
However, the amplitude of the observed spectrum of clusters is about $70$\% 
lower than predicted in these models by using approximation (29). We found a 
similar effect in \S 3 by comparing the analytic and numerical results. It is 
possible that the linear approximation (29) somewhat overestimates the power
spectrum of clusters due to dynamical effects that are not taken into 
account in this approximation. Therefore, the power spectrum of
clusters can be expressed as
$$
P^s_{cl}(k)=\, F \,
\left[1 + {2 f(\Omega_0) \over 3 b_{cl}} + {f^2(\Omega_0) \over
5 b_{cl}^2} \right] b_{cl}^2 \, P(k),
\eqno(32)
$$
where the factor $F=0.7-1.0$. For $F=0.7$, the inflationary models
studied in Figure~5a, are consistent with the power spectrum of the
Abell-ACO clusters at a confidence level of $>90$\% (based on a $\chi^2$ 
test at $k=0.03-0.1h$ Mpc$^{-1}$). Figure~5b demonstrates the power 
spectrum of the model clusters and the APM clusters with a mean 
separation $d_{cl}=31h^{-1}$ Mpc. We have studied the $\Lambda$CDM model 
with $p=1.25$. For $F=0.7$, the cluster power spectrum in this model is 
consistent with the observed power spectrum of the APM clusters at a 
confidence level of $>50$\%. (For $F=0.65$, at a confidence level of 
$>90$\%).

Available data are insufficient to rule out any of the models studied
in Figure~5. Note that the power spectra of clusters predicted in the models
with $p<1$ and $p>1$ are rather different on large scales, where
$k<0.05h$ Mpc$^{-1}$. Therefore, accurate measurements of the power spectrum
of clusters and galaxies on these scales can serve as a discriminating
test for these interesting models.

\section{SUMMARY AND CONCLUSIONS}

Because of the low density of rich clusters of galaxies and their
enhanced clustering strength compared to galaxies, rich clusters
provide a powerful probe of large-scale structure in the Universe.
In this paper, we have examined the power spectrum of clusters in the
Press-Schechter (PS) theory and in N-body simulations to see how the power
spectrum of clusters is related to the power spectrum of matter density
fluctuations in the Universe. We have derived an analytical expression to
determine the correlation function of clusters for a given number
density of clusters in Lagrangian space and have examined this
expression in the linear approximation. In order to
study the power spectrum of clusters in redshift space, we used the
approximation derived by Kaiser (1987).

For testing our analytic results, we used N-body simulations with
realistic initial power spectra of density fluctuations. We
investigated the power spectrum of clusters with a mean separation
$d_{cl}=30-40h^{-1}$ Mpc. The numerical results showed that we can use the
linear PS approximation to predict the power spectrum of clusters in 
real space and in redshift space at wavenumbers $k<0.1h^{-1}$ Mpc.
On smaller scales non-linear processes that are not taken into
account in the linear PS approximation influence the results.

We also used the PS approximation to analyze the redshift-space power
spectrum of clusters in $\Lambda$CDM models. We investigated inflationary
models with a scale-invariant initial spectrum of density
fluctuations and with a steplike initial spectrum derived by
Starobinsky (1992). The results were compared with
observations. We found that the power spectrum of clusters in the
$\Lambda$CDM models with a scale-invariant initial spectrum is not consistent 
with the observed spectra of the APM and Abell clusters. We investigated also
inflationary models with a steplike initial power spectrum (30-31)
that depends on two parameters $k_0$ and $p$. The parameter $k_0$ 
determines the location of the step and the parameter $p$ - the shape of 
the initial spectrum. For the models with $p<1$ and $p>1$, the step 
parameter was chosen to be $k_0=0.016h$ Mpc$^{-1}$ and 
$k_0=0.03h$ Mpc$^{-1}$, respectively. We found that the power spectrum of 
clusters in these $\Lambda$CDM models is in good agreement with the 
observed power spectrum of the Abell-ACO clusters, if the initial 
parameter $p$ is in the range $p=0.6-0.8$. To describe the power spectrum of 
the APM clusters, we can use the $\Lambda$CDM model with parameter
$p=1.25$.

\acknowledgements

We thank J. Einasto, A. Kashlinsky, S. Matarrese, H. Mo and E. Saar for
useful discussions. This work has been supported by the ESF grant 97-2645.

\newpage
\begin{center}
FIGURE CAPTIONS
\end{center}

\figcaption[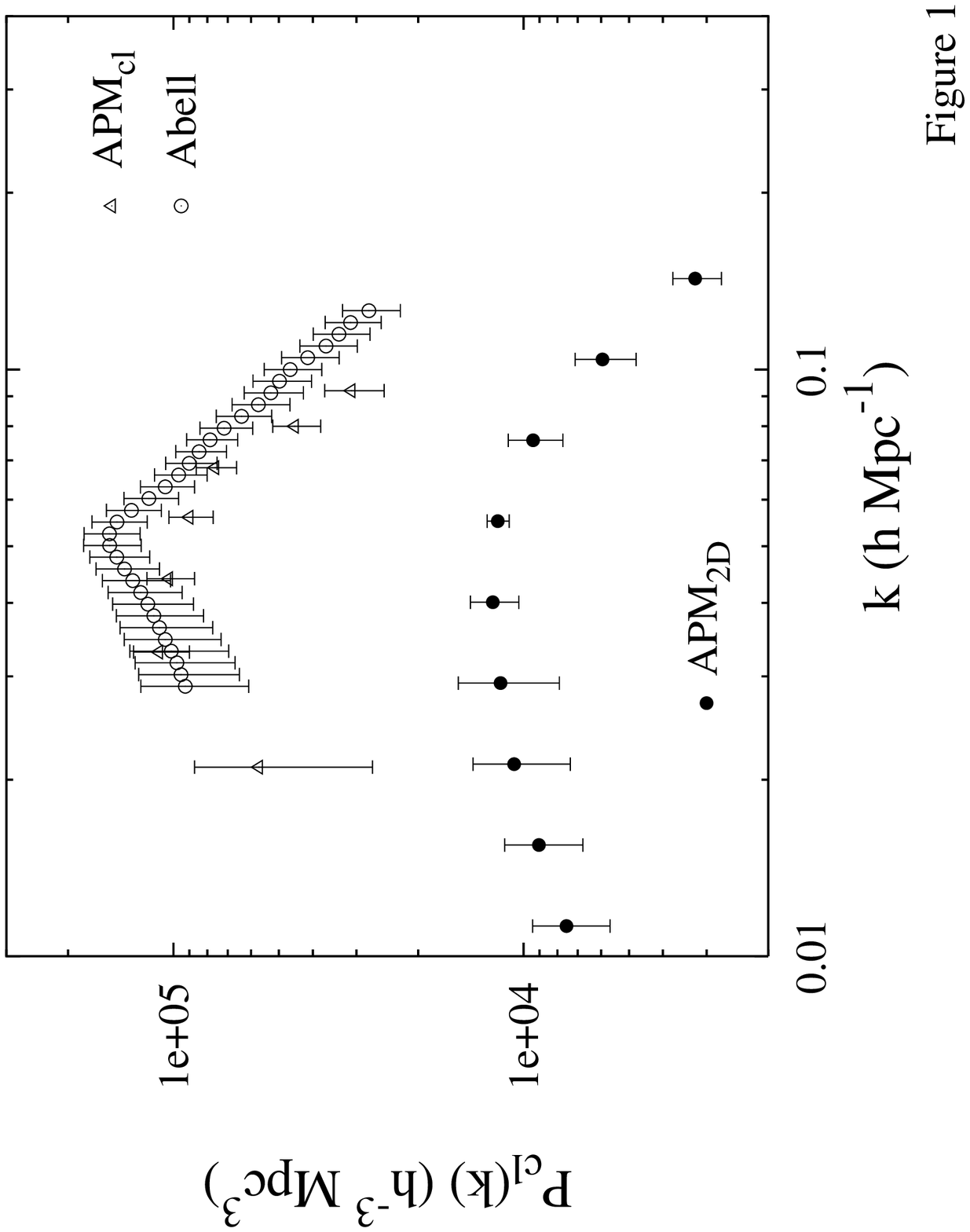]{The power spectrum of the distribution of clusters
of galaxies. Open circles show the power spectrum of the distribution of 
the Abell-ACO clusters and open triangles represent the power spectrum 
of the APM clusters. Filled circles show the power spectrum of the galaxy 
distribution in the APM survey.}

\figcaption[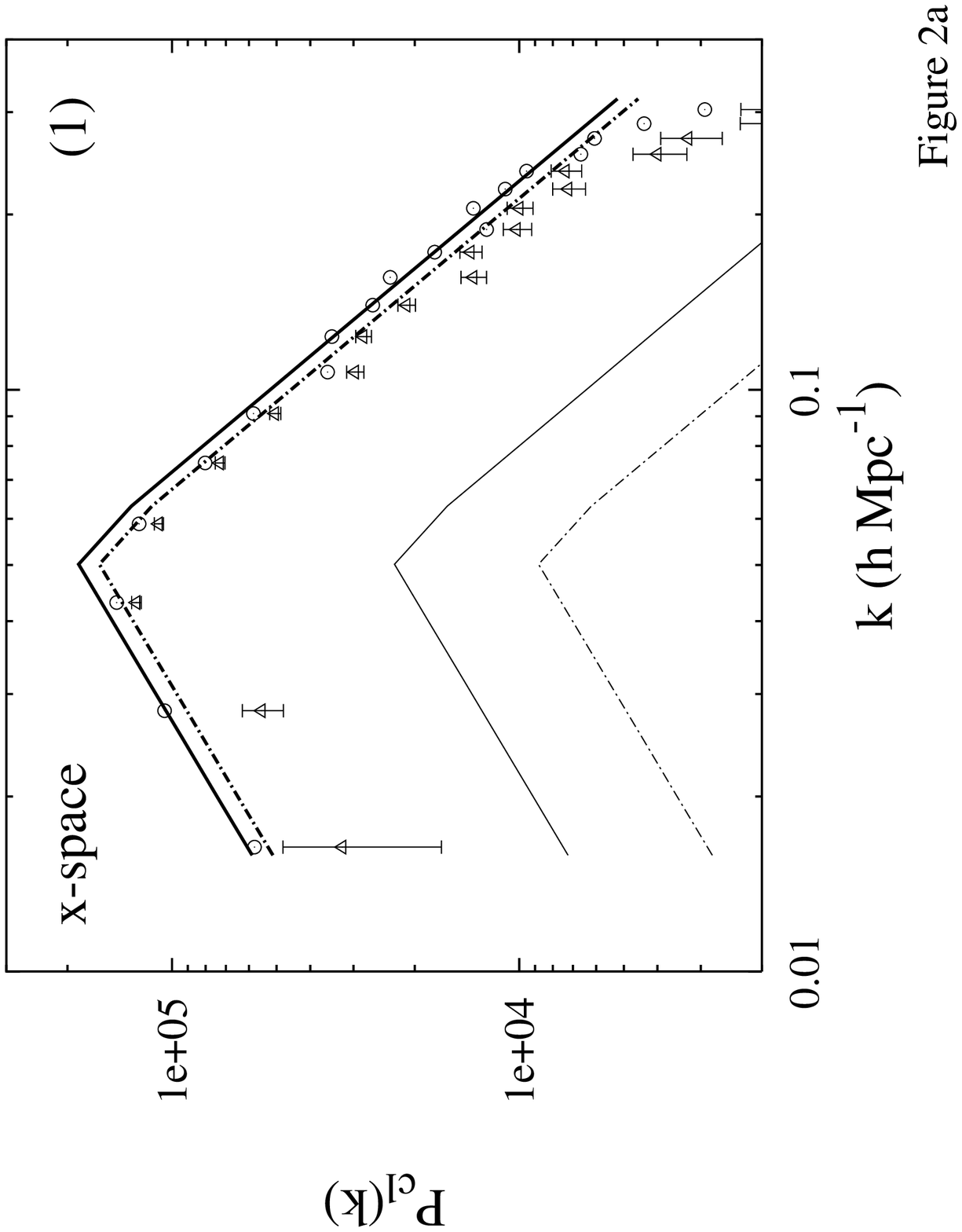,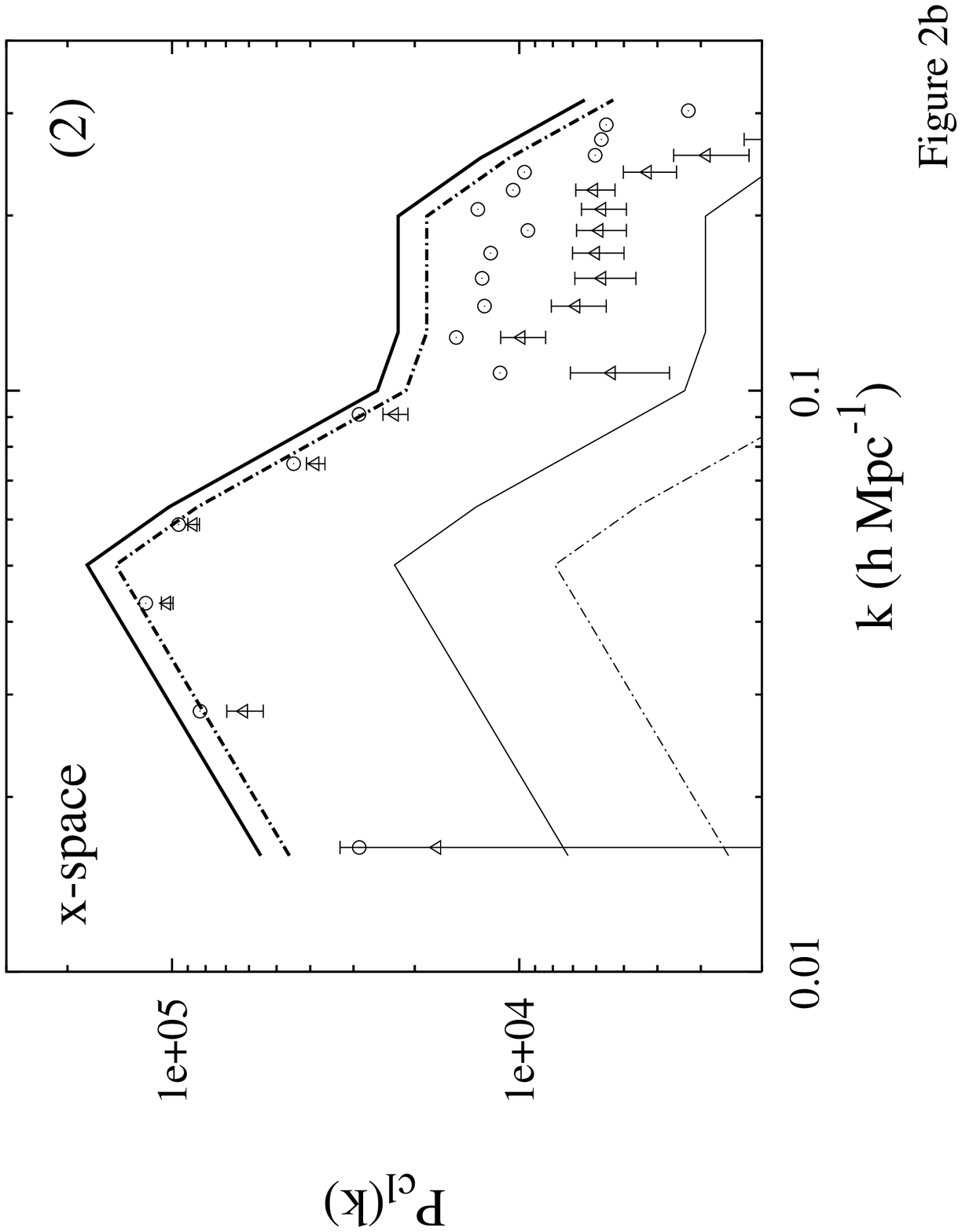]{The power spectrum of the distribution
of clusters in real space. The heavy lines show the power spectra of
clusters determined by approximation (27) and the light lines the
corresponding linear power spectra of matter density fluctuations.
Panel (a) shows the power spectrum in the model (1) for $\sigma_8=0.5$
(dot-dashed lines)
and for $\sigma_8=0.8$ (solid lines). Open circles and triangles
show the power spectra of clusters in N-body simulations for $\sigma_8=0.5$
and $\sigma_8=0.8$, respectively. Panel (b) shows the power spectrum in
the model (2) for $\sigma_8=0.5$ (dot-dashed lines) and for
$\sigma_8=0.84$ (solid lines). Open circles and triangles show the
numerical results for $\sigma_8=0.5$ and $\sigma_8=0.84$, respectively.
Error bars in (a) and (b) denote Poisson errors. The power spectrum is shown
for the clusters with a mean separation $d_{cl}=30h^{-1}$ Mpc.}

\figcaption[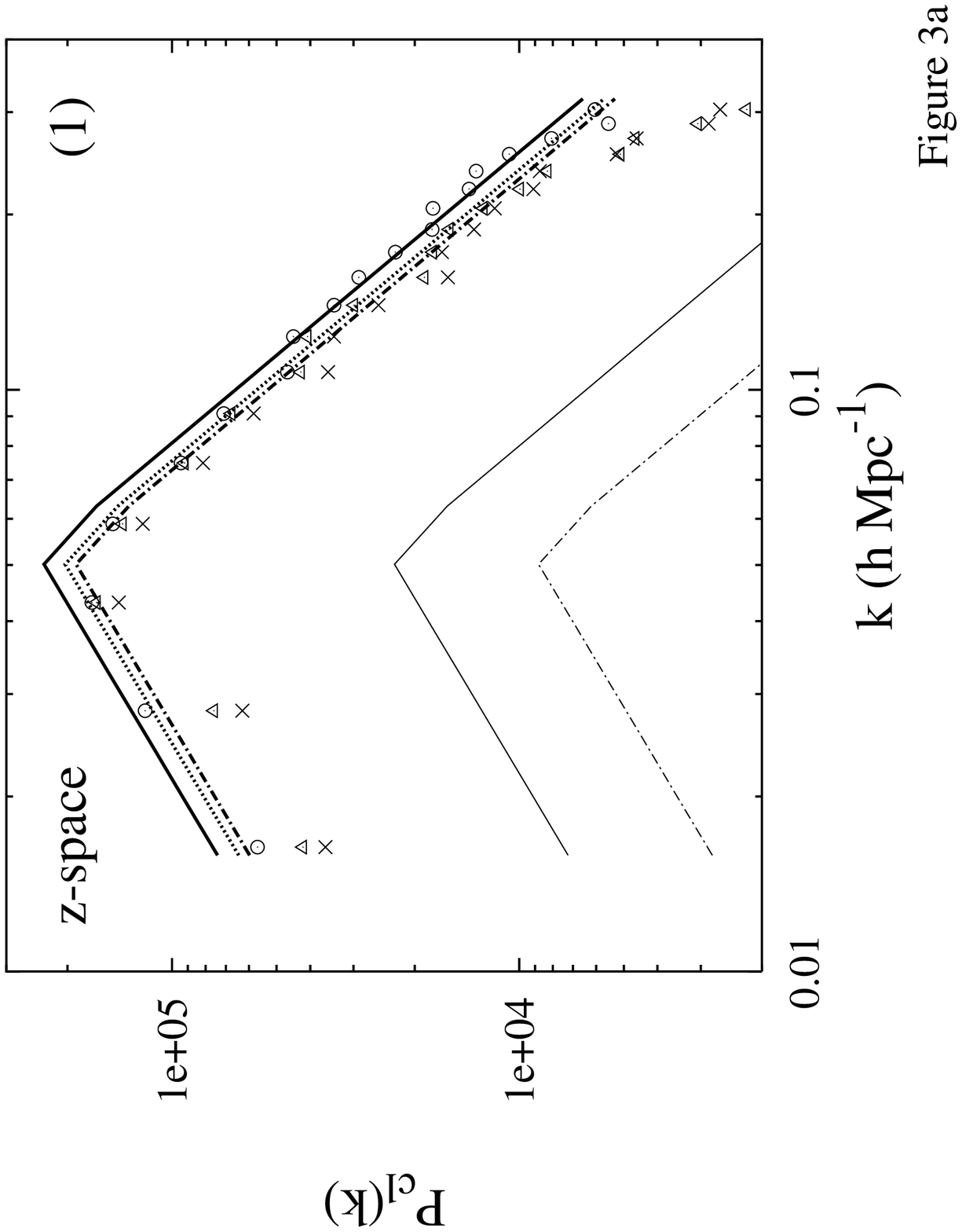,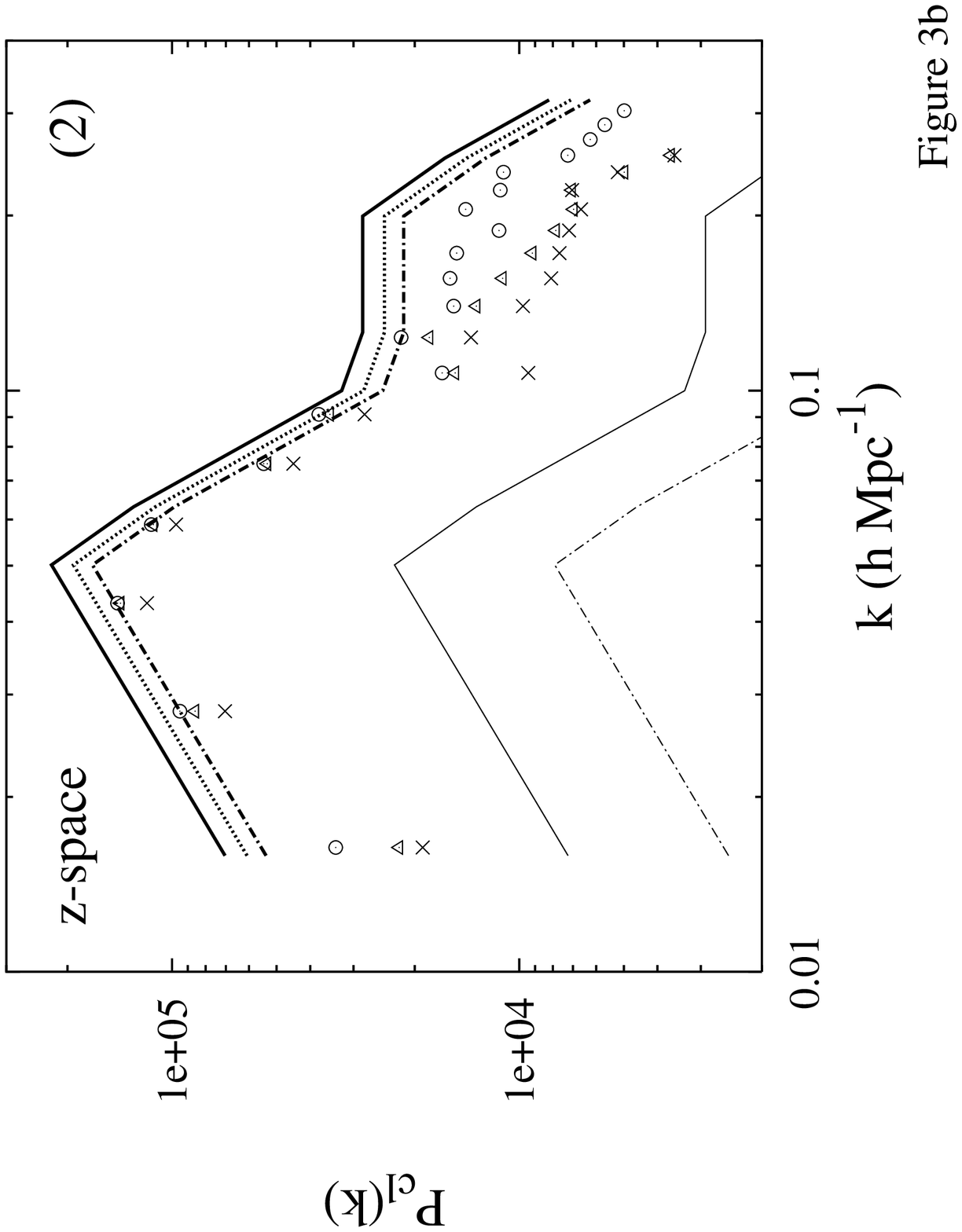]{The power spectrum of the distribution
of clusters in redshift space. Panel (a) shows the results in the model
(1) and panel (b) in the model (2). The heavy lines show the power spectra
of clusters determined by approximation (29) and the light lines the
corresponding linear power spectra of matter fluctuations. Symbols
describe the power spectra of clusters in the simulations.
In the model (1), we studied clusters for $\sigma_8=0.5$, $\Omega=1$
(dot-dashed lines, open circles), for $\sigma_8=0.8$, $\Omega_0=0.2$
(dotted lines, crosses) and for $\sigma_8=0.8$, $\Omega=1$
(solid lines, open triangles). In the model (2), clusters were studied for
$\sigma_8=0.5$, $\Omega=1$ (dot-dashed lines, open circles), for
$\sigma_8=0.84$, $\Omega_0=0.2$ (dotted lines, crosses) and for
$\sigma_8=0.84$, $\Omega=1$ (solid lines, open triangles).
The power spectrum is shown for the clusters with a mean separation
$d_{cl}=30h^{-1}$ Mpc.}

\figcaption[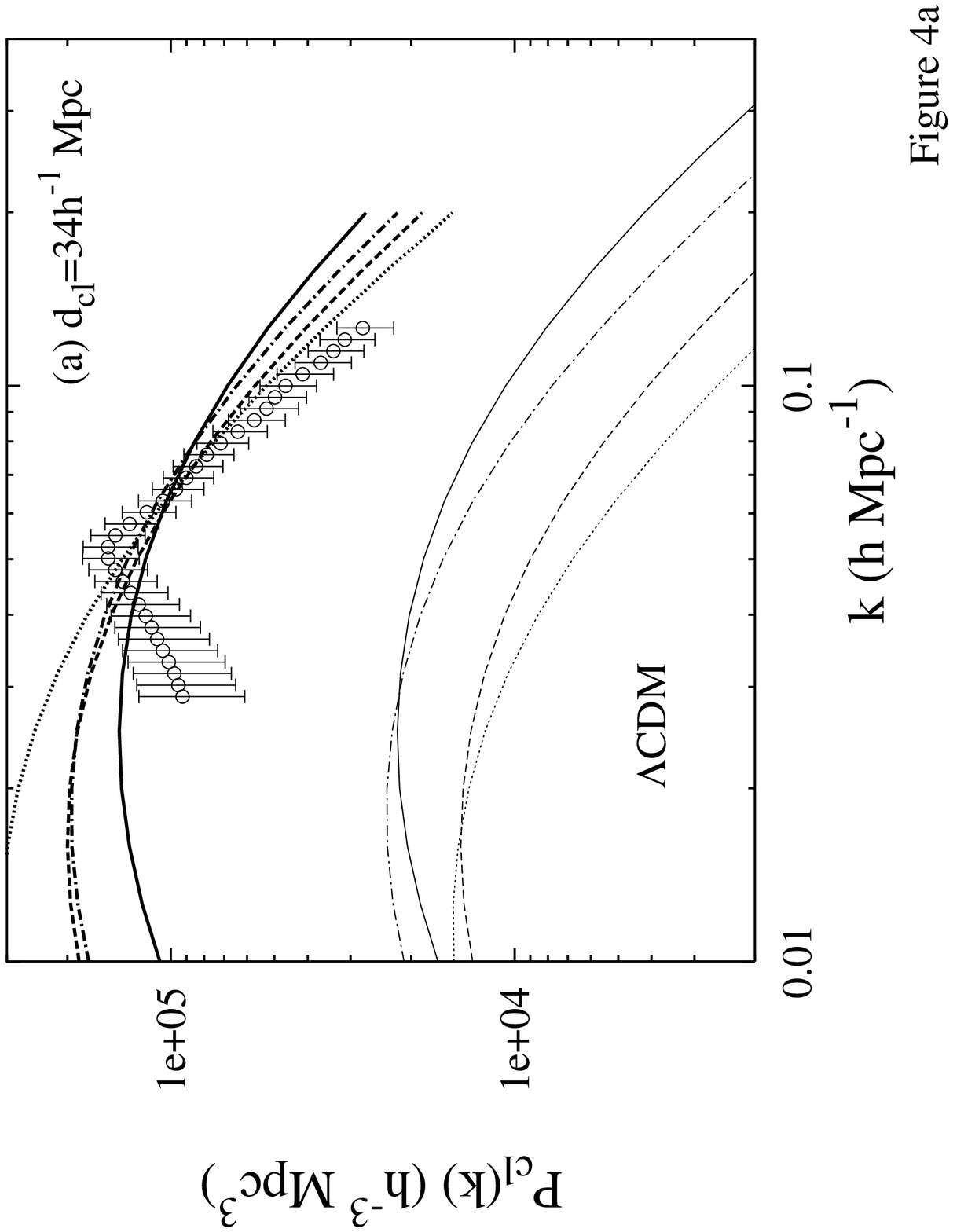,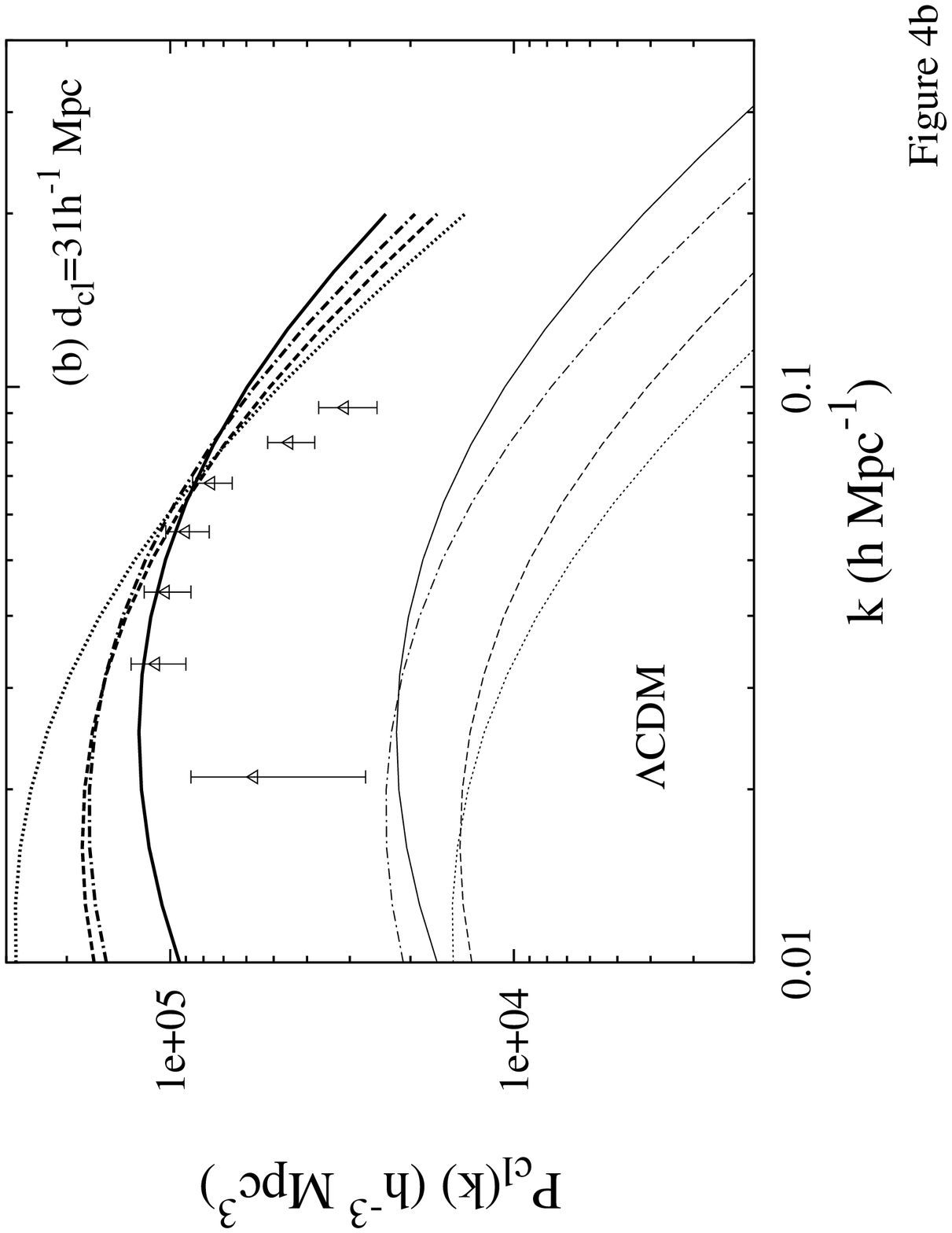]{ The redshift-space power spectrum of
clusters in the $\Lambda$CDM models with $\Omega_0=0.4$, $h=0.7$ (solid
lines), $\Omega_0=0.3$, $h=0.7$ (dot-dashed lines),
$\Omega_0=0.4$, $h=0.5$ (dashed lines) and $\Omega_0=0.3$, $h=0.5$
(dotted lines). The initial spectrum is assumed to be scale-invariant
($P_{in}(k) \propto k$). The heavy lines show the power spectra
of clusters determined by approximation (29) and the light lines the 
corresponding linear power spectra of matter fluctuations. (a) The power 
spectrum of the model clusters and the Abell-ACO clusters (open
circles) with a mean separation $d_{cl}=34h^{-1}$ Mpc. (b) The power 
spectrum of the model clusters and the APM clusters (open triangles) 
with a separation $d_{cl}=31h^{-1}$ Mpc.}

\figcaption[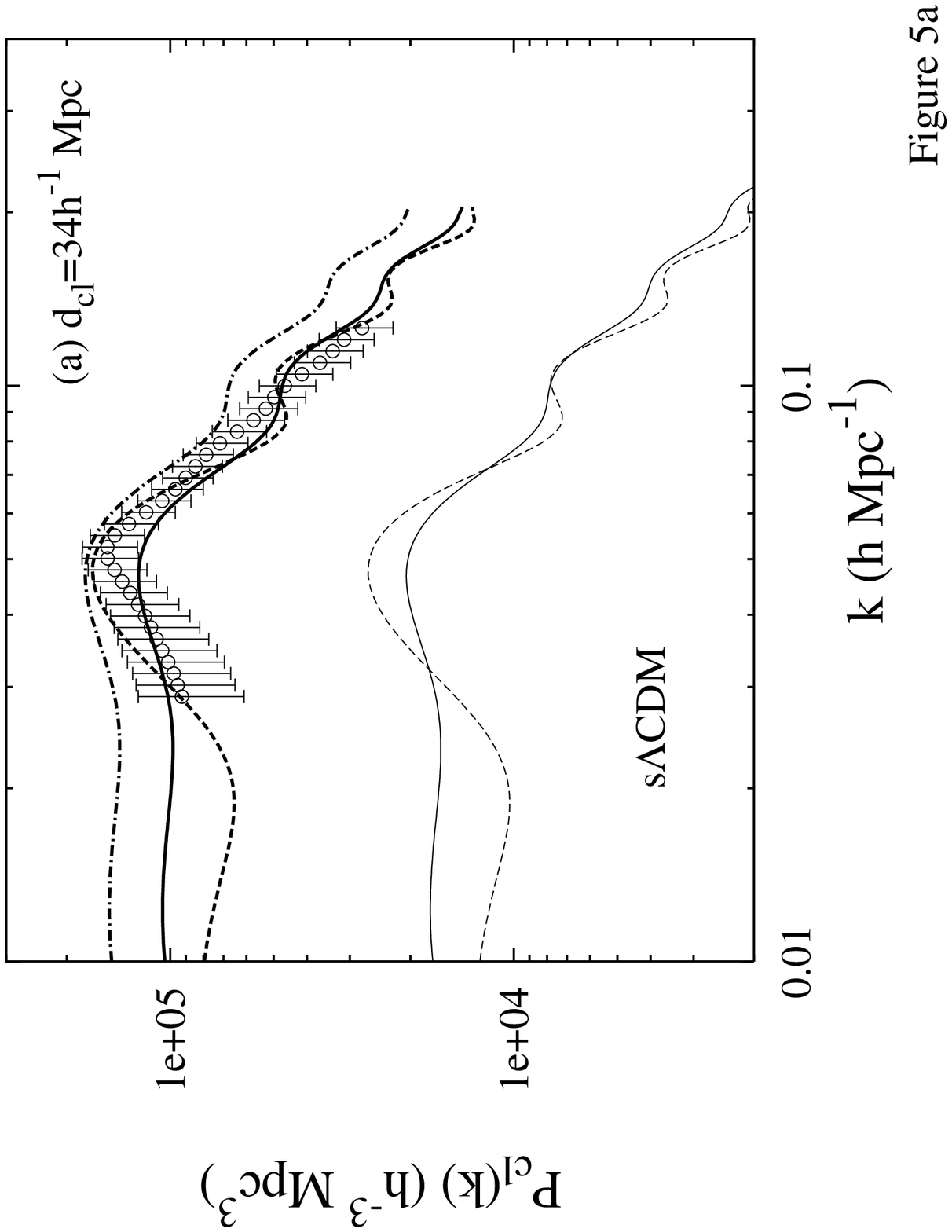,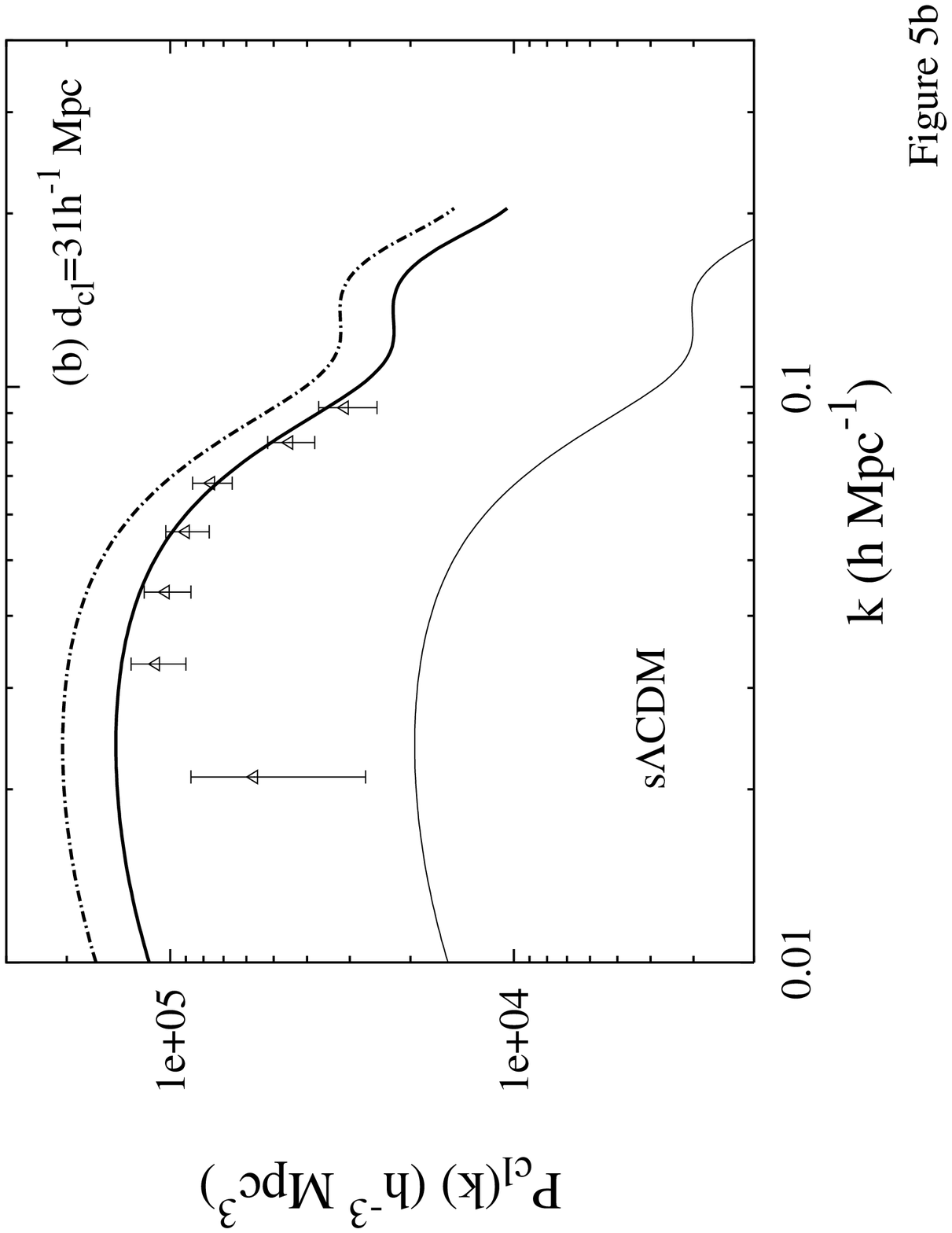]{ The redshift-space power spectrum of
clusters in the $\Lambda$CDM models with a steplike initial power spectrum.
The heavy lines show the power spectra of clusters determined by
approximation (32) and the light lines the corresponding
linear power spectra of matter fluctuations. (a) The power spectrum of 
the model clusters and the Abell-ACO clusters (open circles) with a separation 
$d_{cl}=34h^{-1}$ Mpc. We have studied the models with $\Omega_0=0.3$, 
$h=0.6$, $p=0.8$ (solid lines) and $\Omega_0=0.3$, $h=0.5$, $p=0.6$
(dashed lines). The cluster power spectrum is determined for $F=0.7$.
The dot-dashed line shows the cluster power spectrum in the first model
for $F=1.0$. (b) The power spectrum of the model 
clusters and the APM clusters (open triangles) with a mean separation 
$d_{cl}=31h^{-1}$ Mpc. We have studied the model with 
$\Omega_0=0.4$, $h=0.6$ and $p=1.25$. The power spectum of clusters is
determined for $F=0.7$ (solid line) and $F=1.0$ (dot-dashed line).}

\end{document}